\documentclass{PoS}
\usepackage[sort&compress,numbers,merge]{natbib}
\usepackage{natbib}

\title{On the continuum limit of Landau gauge gluon \\
and ghost propagators in $SU(2)$ lattice gauge gluodynamics}

\ShortTitle{Continuum limit of propagators}

\author{\speaker{Igor Bogolubsky}\\
      Joint Institute for Nuclear Research, LIT, 141980 Dubna, Russia\\
        E-mail: \email{bogolubs@jinr.ru}}

\author{Ernst-Michael Ilgenfritz\\
 Joint Institute for Nuclear Research, VBLHEP, 141980 Dubna, Russia\\
        E-mail: \email{michael.ilgenfritz@lhep.jinr.ru}}

\author{Michael M\"uller-Preussker\\
 Humboldt-Universit\"at zu Berlin, Institut f\"ur Physik, 12489 Berlin, Germany\\
        E-mail: \email{mmp@physik.hu-berlin.de}}

\author{Andre Sternbeck\\
Universit\"at Regensburg, Institut f\"ur Theoretische Physik, 93040 Regensburg, Germany \\
        E-mail: \email{andre.sternbeck@physik.uni-regensburg.de}}

\abstract{We continue the systematic computation of Landau gauge
gluon and ghost propagators of $SU(2)$ gluodynamics using a sequence of 
increasing lattice sizes $L^4$ up to $L=112$ with corresponding $\beta$-values
chosen to keep the linear physical size $a(\beta) L \simeq 9.6 \mathrm{~fm} $
fixed. To extremize the Landau gauge functional we employ simulated annealing
combined with subsequent overrelaxation. Renormalizing the propagators at 
momentum  $\mu= 2.2$ GeV we observe quite strong lattice artifacts
for the gluon propagator as well as for the ghost dressing function within the 
momentum region $q < 1.0 $ GeV. The dependence on the lattice spacing for the 
gluon propagator at lowest accessible physical momentum values does not yet
allow a simple extrapolation to the continuum limit. On the contrary, the 
running coupling derived from the bare dressing functions seems less affected 
by lattice artifacts.}

\FullConference{Xth Quark Confinement and the Hadron Spectrum,\\
        October 8-12, 2012\\
        TUM Campus Garching, Munich, Germany}

\begin{document}
The aim of the reported study is to continue the systematic investigation 
of $SU(2)$ gluon and ghost Landau gauge propagators on large lattices
\cite{Bogolubsky:2009qb}
in order to receive information from first principles on the behavior of 
these propagators and of the running coupling in the continuum limit for 
all momenta $q$ including the infrared (IR) region. As in previous investigations 
\cite{Bakeev:2003rr,*Bogolubsky:2005wf,*Bogolubsky:2007bw,*Bornyakov:2009ug}, 
for solving the Gribov problem we assume the Landau gauge functional 
to be driven as close as possible to its global extremum. 
Employing the standard Wilson plaquette action we have studied gluon ($D$) 
and ghost ($G$) propagators for lattice sizes $L^4$ with run parameters 
collected in Table~\ref{tab:table1}.
The gluon propagator was computed using $N_{MC}(\beta, L)$ independent
Monte Carlo (MC) configurations generated with the given set of
parameters, while the ghost propagator was calculated only on
a subset of $N_{ghost}(\beta,L)$ MC configurations.
%-------------------------------------------------------------------
\begin{table}
\begin{center}
\begin{tabular}{cccccccc}
\hline
 $\beta $&2.3 &2.3 &2.3 &2.3 &2.4 &2.45 & 2.5 \\
\hline
 $L$&40
 &56 &80 &112&80 &96 &112 \\
\hline
 $N_{MC}$ &45 &187 &78 &173 &314&
 333&477 \\
 $N_{ghost}$ & - &24 & - &- &26& 1 &- \\
$\tilde{Z}(\mu=2.2$~GeV)&  -  & 0.414(3) & - & - & 0.445(4) & 0.452(2) & 0.460(5)\\
\hline
\end{tabular}
\end{center}
\caption{Parameters of the main simulations considered here. See also the text.}
\label{tab:table1}
\end{table}
%--------------------------------------------------------------------------
We used the same procedures for Landau gauge fixing and computation
of propagators as described in \cite{Bogolubsky:2009qb}, namely, we employed
very long simulated annealing (SA) runs followed by overrelaxation (OR) 
to obtain gauge copies with a gauge fixing functional close to its global 
extremum for each MC configuration.
%--------------------------------------------------------------------------
\begin{figure}
\begin{center}
\mbox{
\includegraphics[height=7cm,angle=270]{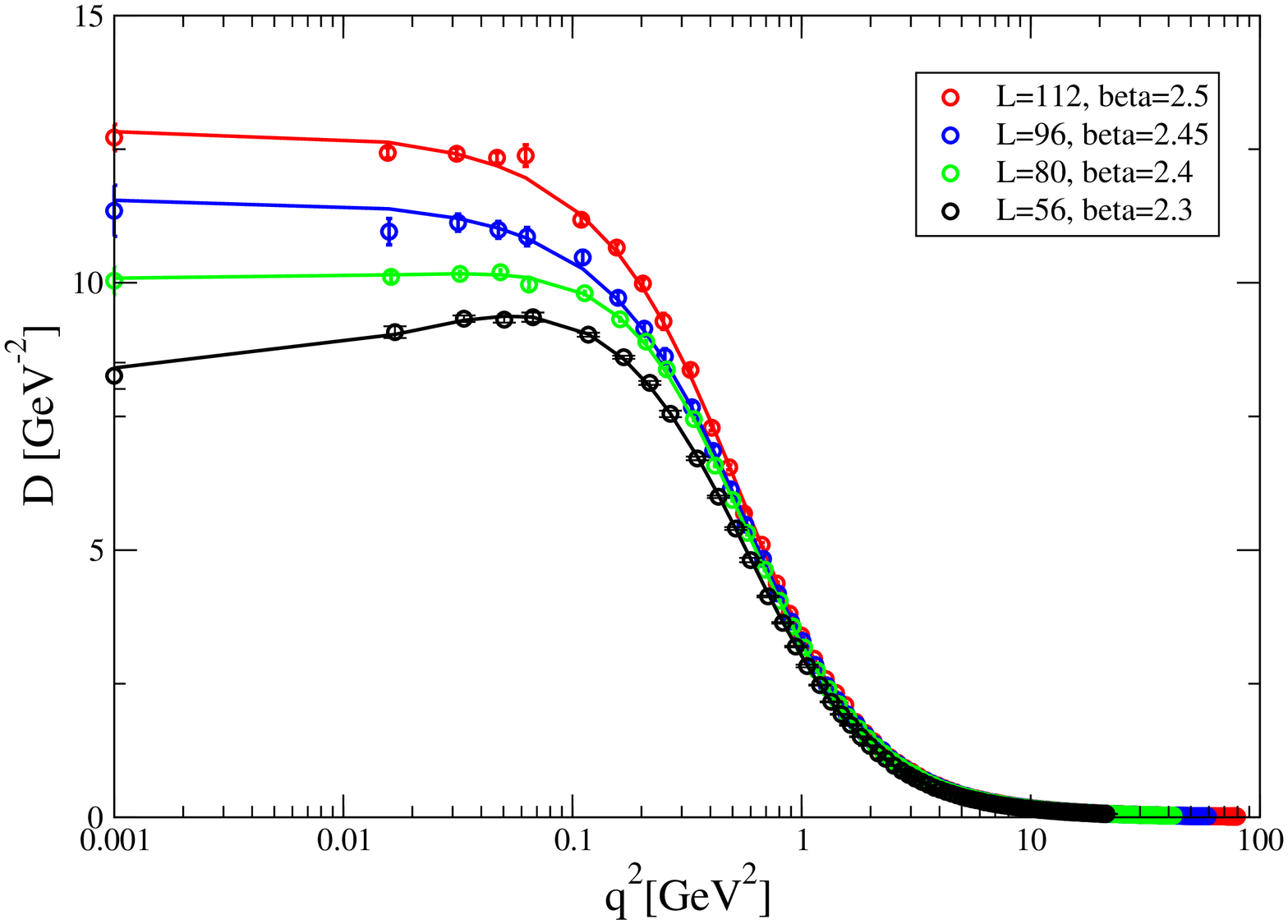}
\hspace{5ex}
\includegraphics[height=7cm,angle=270]{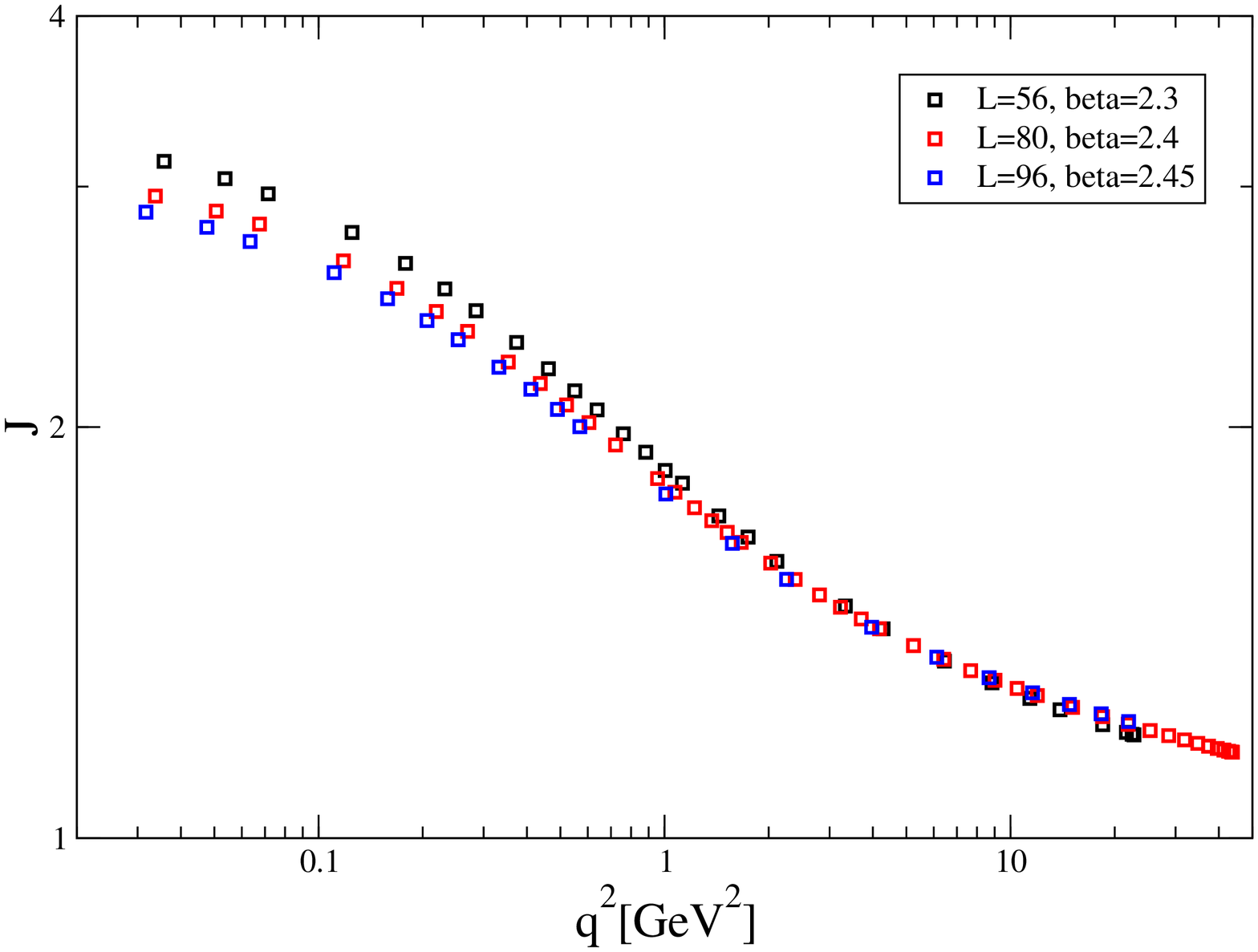}
}
\caption{{\bf~Left}:~the unrenormalized gluon propagator $D(q^2)$ and its
fits according to Eq.~(1); the data points drawn at $q^2 = 0.001$
 represent the zero-momentum gluon propagator $D(0)$ values.
{\bf~Right}: the ghost dressing function $J(q^2)$. Both are shown 
for approximately equal physical volume ($~(aL)^4 \simeq (9.6 \mathrm{fm})^4$) 
but different $\beta$-values, i.e. discretization scales $a$.}
\label{fig:fig1}
\end{center}
\end{figure}
%-----------------------------------------------------------------------
In Fig.~\ref{fig:fig1} we show the results for the bare gluon propagator
(left) and the bare ghost dressing function (right) for fixed physical
volume $(aL)^4$ but varying lattice scale $a$. For the gluon propagator
we have drawn also curves obtained from fits with the 6-parameter formula 
proposed in \cite{Cucchieri:2011ig}
\begin{equation}
\label{eq:fit}
  D(q)=C\frac{q^4+A^2q^2+B}{q^6+Eq^4+Fq^2+G^2}\;.
\end{equation}
We found the resulting fit curves nicely to capture the IR turnover 
of the gluon propagator. The $\chi^2/dof$ values are close to unity in 
most cases (see Table~\ref{tab:table2}).
%------------------------------------------------------------------------
\begin{table}
\begin{center}
\begin{tabular}{cccccccc}
\hline
 $\beta$&$ C $ & $A$ & $B$ & $E$ & $F$ & $G$ &
  $\chi^2/dof$ \\
\hline
 $2.3$&1.164(3)&1.52(1)
 &0.11(3)&0.68(2)&0.502(4)& 0.126(15)&1.03\\
\hline
 $2.4$  &1.195(2) &1.68(2)&1.1(3)& 1.13(10)&0.60(3)&
 0.36(5)&2.73 \\
\hline
 $2.45$ &1.148(11)&3.9(5)&99(34)
 &22(7)&12(4)&3.1(5)&0.28 \\
\hline
 $2.5$&1.127(8)&5.47(34)&210(33)&
 45(6)&25(4)&4.3(3)&1.20 \\
\hline
\end{tabular}
\end{center}
\caption{Results of the 6-parameter fits of the unrenormalized gluon
propagator for various $\beta$ corresponding to lattices sizes $56^4$,
$80^4$, $96^4$ and $112^4$.}
\label{tab:table2}
\end{table}
%---------------------------------------------------------------------
To obtain the renormalized gluon propagator 
$D_{ren}(q,\mu)=\tilde{Z}(a,\mu)D(q,a))$ we apply the normalization
condition $D_{ren}(\mu,\mu)=1/\mu^2$. Since the fit formula  
Eq.~(\ref{eq:fit}) nicely works throughout the whole momentum region
we can use it to carry out the renormalization at any $\mu$. For 
$\mu=2.2$~GeV the renormalization factors $\tilde{Z}$ are 
collected in Table \ref{tab:table1}. Their values do not vary strongly
which means that the bare dressing functions for the different 
$\beta$-values approximately overlap at the given $\mu$-value.
From Fig. \ref{fig:fig1} we can then conclude also for the renormalized 
gluon propagator $D_{ren}(q^2)$ and the renormalized ghost dressing 
function $J_{ren}(q^2)$ found for various lattice spacings $a(\beta)$ 
to be compatible with the so-called {\it decoupling solution} of 
Dyson-Schwinger or functional renormalization group equations 
(see \cite{Fischer:2008uz}).   
The numerical values, however, of $D_{ren}(q^2)$ and $J_{ren}(q^2)$
in the limit $q \to 0$ appear to be $\beta$- or $a$-dependent. 
From such plots one can see that the convergence of the renormalized
lattice propagators / dressing functions in the deep IR momentum 
range to the respective continuum counterpart,
that should be observed for decreasing $a(\beta)$, is rather slow.
For their direct numerical study near the continuum limit
one has to use rather large $\beta$-values which consequently requires
simulations on unrealistically huge lattices, which are not
accessible today even on most powerful parallel supercomputers.
Instead, we can try to make contact with the continuum limit by
extrapolating $D_{ren}(q^2,a)$ to the zero-$a$ limit as done e.g. 
in \cite{Aouane:2011fv} for $SU(3)$ and non-zero temperature.
In Fig.~\ref{fig:Dren_vs_a} (left) we plot the $a$-dependence of 
lattice $D_{ren}(q^2)$ for several selected values of $q^2$.
We see the lower the momentum is the less well-defind the convergence 
for $a \to 0$ becomes. For getting reliable numerical values of $SU(2)$ 
gluon and ghost propagators in the continuum limit more work is needed.
Although the ghost dressing function $J(q^2)$ has been computed only for 
a subset of MC configurations (see Table~\ref{tab:table2}), it provided 
useful quantitative information, see Fig.~\ref{fig:fig1} (right).
Even a single MC configuration as for $L=96$ already seems to yield 
a first estimate (a fast decrease of the statistical fluctuations of 
$J(q^2)$ with increasing $L$ was first observed for the $SU(3)$ case in 
\cite{Bogolubsky:2007ud}).
Our analysis shows in the deep IR region that, while $D_{ren}(q^2)$  
increases with $\beta$, $J_{ren}(q^2)$ decreases. More details will be 
published elsewhere.
%--------------------------------------------------------------------
\begin{center}
\begin{figure}
\mbox{
\includegraphics[height=7cm,angle=270]{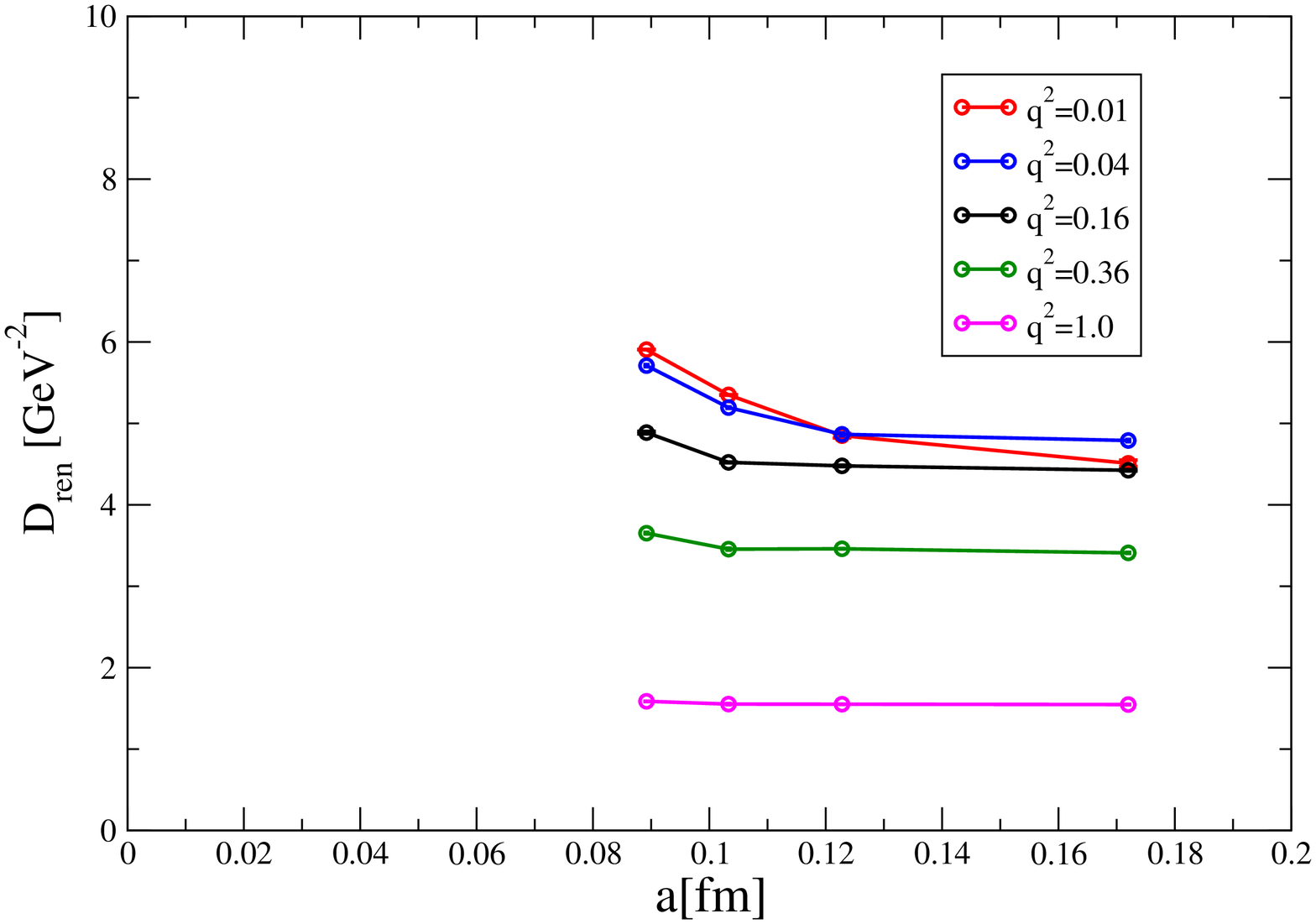}
\hspace{5ex}
\includegraphics[height=7cm,angle=270]{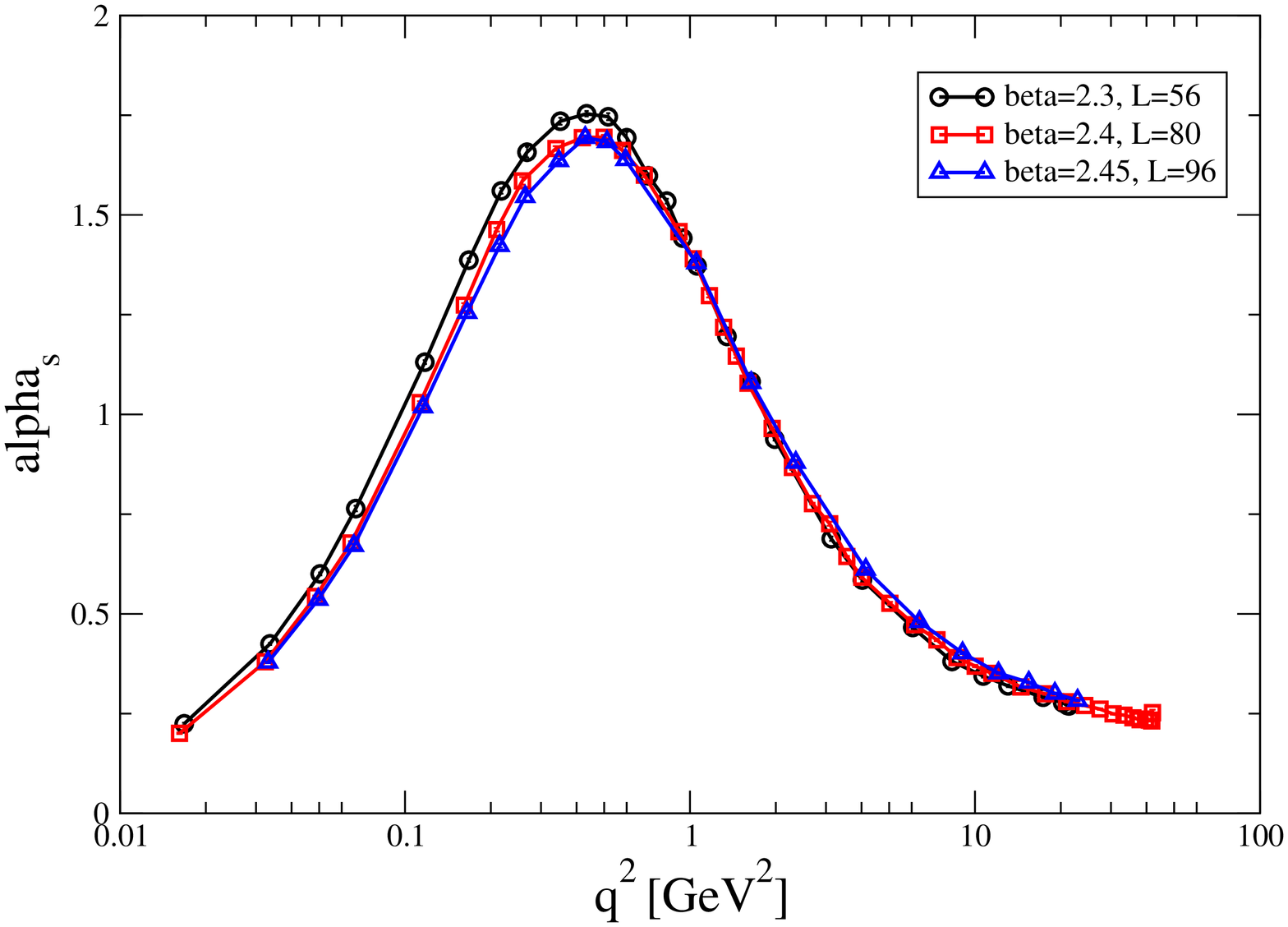}
 }
\caption{{\bf~Left:} Renormalized gluon propagator $D_{ren}(q^2)$ 
at $\mu=2.2$~GeV versus lattice spacing $a$ for various physical momenta
$q^2$  (values indicated in units GeV$^2$).
{\bf Right:} Running coupling for $\beta=2.3$, $\beta=2.4$ and $\beta=2.45$}
\label{fig:Dren_vs_a}
\end{figure}
\end{center}
%---------------------------------------------------------------------

\begin{center}
\begin{figure}
\mbox{
\includegraphics[height=7cm,angle=270]{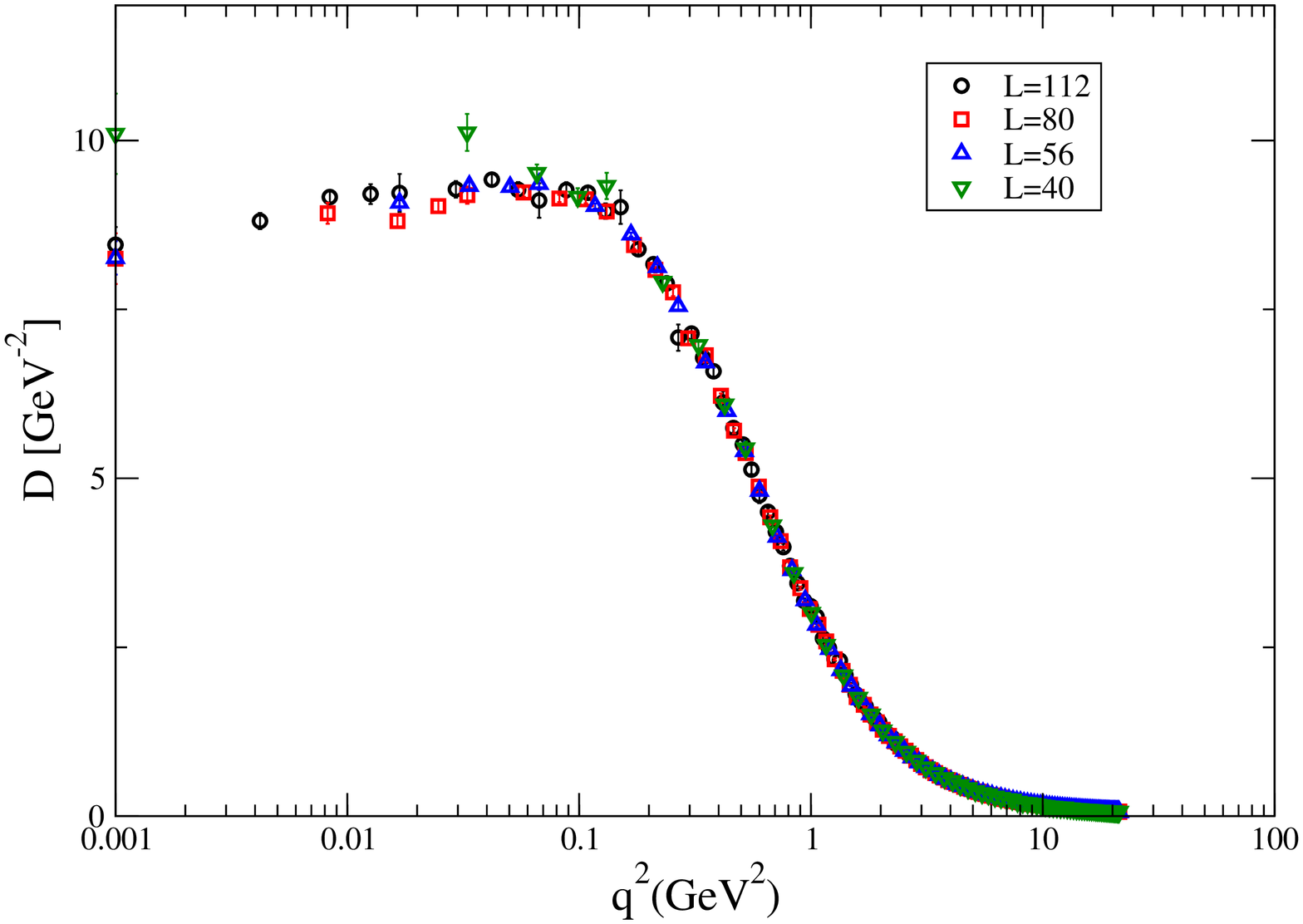}}
\hspace{5ex}
\includegraphics[height=7cm,angle=270]{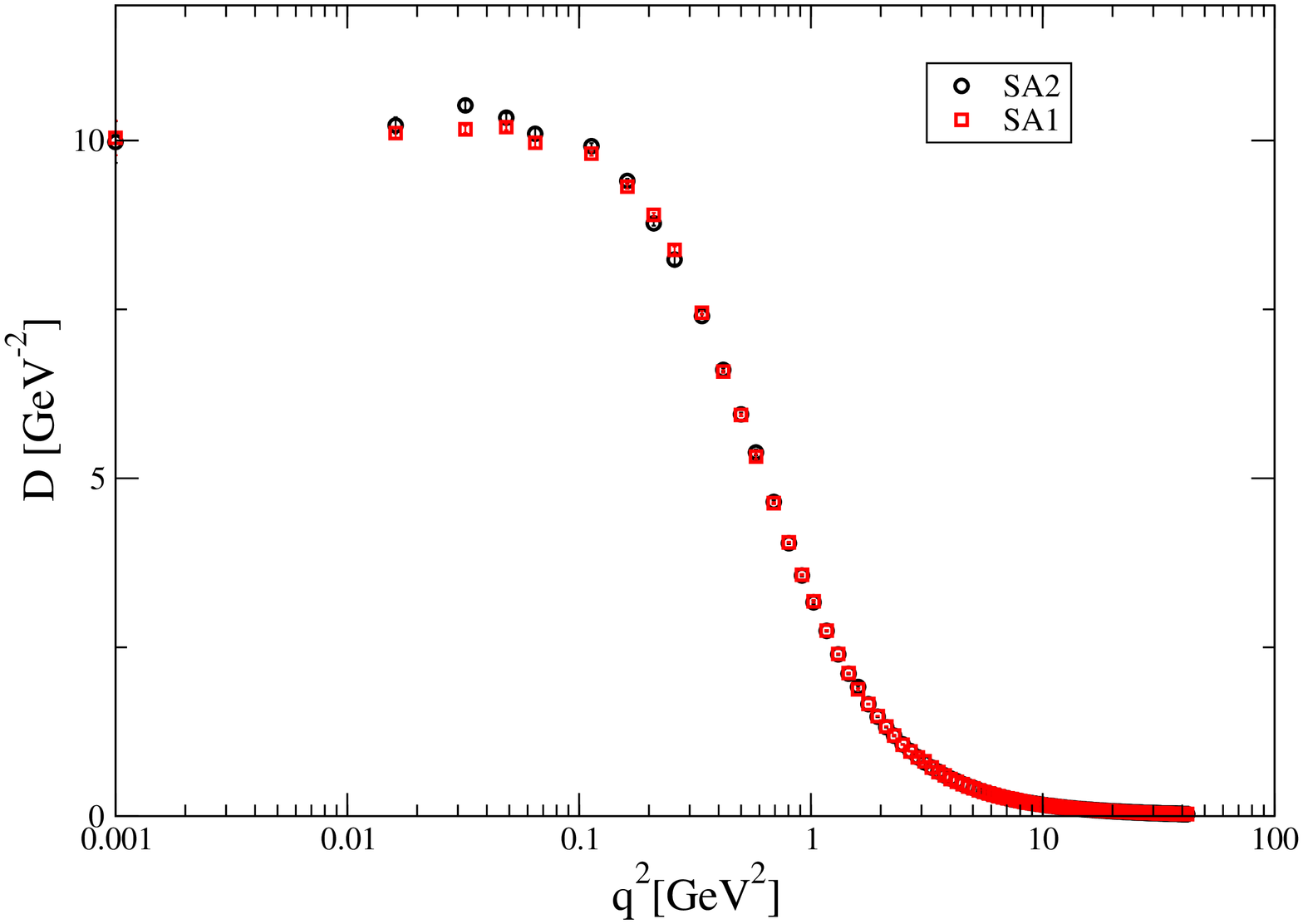}
\caption{Check of systematic errors. {\bf~Left:} The unrenormalized
$D(q^2)$ computed for $\beta=2.3$ and various
$L$. {\bf Right:} $D(q^2)$ for 2 different SA schedules at $\beta=2.4$
and $L=80$.}
\label{fig:FVE}
\end{figure}
\end{center}

We have checked whether the differences of propagator
values in the deep IR could be compensated by other systematic effects.
From Fig.~\ref{fig:FVE} (left) one can see that finite-volume effects 
are small if the linear physical size is $a(\beta) L \simeq 9.6 $ fm or 
even larger. What concerns Gribov copy artifacts at $\beta=2.4$ and $L=80$
we have compared the results of two sets of SA+OR gauge fixing simulations: 
(i) one gauge copy fixing with 9600 SA sweeps ("SA1 schedule") $(N_{MC}=314)$
and (ii) "best of two copies" gauge fixing with 12000 SA sweeps each
("SA2 schedule") $(N_{MC}=187)$. For details see,
e.g., Ref. \cite{Bogolubsky:2009dc}. SA1 and SA2 results obtained for
unrenormalized gluon propagators are plotted in Fig.~\ref{fig:FVE}
(right). We have found that the differences between these cases
are much smaller than the magnitude of Gribov copy effects 
measured in \cite{Bogolubsky:2009qb} as
difference between results of one-copy SA+OR and one-copy OR
gauge-fixing procedures. Our analysis shows that further
``improvement'' of SA schedules could not change $D_{ren}(q^2)$
essentially and hence noticeable differences of $D_{ren}(q^2)$
values in the deep IR region found for different $\beta$-values certainly
cannot be accounted for by the Gribov copy effect. 
%-------------------------------------------------------------------------
%------------------------------------------------------------------------

With the bare gluon ($Z(q^2)$) and ghost ($J(q^2)$) dressing functions at  
hand one can easily compute the running coupling in the minimal MOM scheme
\cite{vonSmekal:2009ae},
$$ \alpha_s(q^2)~=~\frac{g_0^2}{4 \pi}~J^2(q^2)~Z(q^2).$$
The dependence of the resulting curves on $\beta$ or $a$
turns out to be rather weak even in the deep IR momentum
region (see Fig.~\ref{fig:Dren_vs_a}~(right)), i.e. the lattice
artifacts of the gluon and ghost dressing functions cancel each
other to some extent.

We conclude that naive multiplicative renormalizability for the
$SU(2)$ Landau gauge gluon and ghost propagators gets violated
in the deep IR region. Due to the slow convergence of gluon and 
ghost renormalized propagators their continuum counterparts may 
strongly differ in the deep IR momentum region from what we have 
obtained here in lattice simulations with admissible
values of $\beta=4/g_0^2$. At the same time, the physically 
important renorm-invariant minimal MOM-scheme running coupling 
$\alpha_s(q^2)$ seems to reach a continuum behavior much earlier. 

IB thanks Prof. A.~A.~Slavnov for a useful discussion of the results. 
Simulations have been done on the MVS100K supercomputer of the
Joint Supercomputer Centre (JSCC, Moscow).

\bibliographystyle{apsrev4-1}
%\bibliography{citations.bib}

\begin{thebibliography}{11}%
\makeatletter
\providecommand \@ifxundefined [1]{%
 \@ifx{#1\undefined}
}%
\providecommand \@ifnum [1]{%
 \ifnum #1\expandafter \@firstoftwo
 \else \expandafter \@secondoftwo
 \fi
}%
\providecommand \@ifx [1]{%
 \ifx #1\expandafter \@firstoftwo
 \else \expandafter \@secondoftwo
 \fi
}%
\providecommand \natexlab [1]{#1}%
\providecommand \enquote  [1]{``#1''}%
\providecommand \bibnamefont  [1]{#1}%
\providecommand \bibfnamefont [1]{#1}%
\providecommand \citenamefont [1]{#1}%
\providecommand \href@noop [0]{\@secondoftwo}%
\providecommand \href [0]{\begingroup \@sanitize@url \@href}%
\providecommand \@href[1]{\@@startlink{#1}\@@href}%
\providecommand \@@href[1]{\endgroup#1\@@endlink}%
\providecommand \@sanitize@url [0]{\catcode `\\12\catcode `\$12\catcode
  `\&12\catcode `\#12\catcode `\^12\catcode `\_12\catcode `\%12\relax}%
\providecommand \@@startlink[1]{}%
\providecommand \@@endlink[0]{}%
\providecommand \url  [0]{\begingroup\@sanitize@url \@url }%
\providecommand \@url [1]{\endgroup\@href {#1}{\urlprefix }}%
\providecommand \urlprefix  [0]{URL }%
\providecommand \Eprint [0]{\href }%
\providecommand \doibase [0]{http://dx.doi.org/}%
\providecommand \selectlanguage [0]{\@gobble}%
\providecommand \bibinfo  [0]{\@secondoftwo}%
\providecommand \bibfield  [0]{\@secondoftwo}%
\providecommand \translation [1]{[#1]}%
\providecommand \BibitemOpen [0]{}%
\providecommand \bibitemStop [0]{}%
\providecommand \bibitemNoStop [0]{.\EOS\space}%
\providecommand \EOS [0]{\spacefactor3000\relax}%
\providecommand \BibitemShut  [1]{\csname bibitem#1\endcsname}%
\let\auto@bib@innerbib\@empty
%</preamble>
\bibitem [{\citenamefont {Bogolubsky}\ \emph
  {et~al.}(2009{\natexlab{a}})\citenamefont {Bogolubsky}, \citenamefont
  {Ilgenfritz}, \citenamefont {M{\"u}ller-Preussker},\ and\ \citenamefont
  {Sternbeck}}]{Bogolubsky:2009qb}%
  \BibitemOpen
  \bibfield  {author} {\bibinfo {author} {\bibfnamefont {I.~L.}\ \bibnamefont
  {Bogolubsky}}, \bibinfo {author} {\bibfnamefont {E.-M.}\ \bibnamefont
  {Ilgenfritz}}, \bibinfo {author} {\bibfnamefont {M.}~\bibnamefont
  {M{\"u}ller-Preussker}}, \ and\ \bibinfo {author} {\bibfnamefont
  {A.}~\bibnamefont {Sternbeck}},\ }\href@noop {} {\bibfield  {journal}
  {\bibinfo  {journal} {PoS}\ }\textbf {\bibinfo {volume} {(LAT2009) 237}}
  (\bibinfo {year} {2009}{\natexlab{a}})},\ \Eprint
  {http://arxiv.org/abs/0912.2249} {arXiv:0912.2249 [hep-lat]} \BibitemShut
  {NoStop}%
\bibitem [{\citenamefont {Bakeev}\ \emph {et~al.}(2004)\citenamefont {Bakeev},
  \citenamefont {Ilgenfritz}, \citenamefont {Mitrjushkin},\ and\ \citenamefont
  {M{\"u}ller-Preussker}}]{Bakeev:2003rr}%
  \BibitemOpen
  \bibfield  {author} {\bibinfo {author} {\bibfnamefont {T.~D.}\ \bibnamefont
  {Bakeev}}, \bibinfo {author} {\bibfnamefont {E.-M.}\ \bibnamefont
  {Ilgenfritz}}, \bibinfo {author} {\bibfnamefont {V.~K.}\ \bibnamefont
  {Mitrjushkin}}, \ and\ \bibinfo {author} {\bibfnamefont {M.}~\bibnamefont
  {M{\"u}ller-Preussker}},\ }\href@noop {} {\bibfield  {journal} {\bibinfo
  {journal} {Phys. Rev.}\ }\textbf {\bibinfo {volume} {D69}},\ \bibinfo {pages}
  {074507} (\bibinfo {year} {2004})},\ \Eprint
  {http://arxiv.org/abs/hep-lat/0311041} {hep-lat/0311041} \BibitemShut
  {NoStop}%
%%CITATION = HEP-LAT 0311041;%%
\bibitem [{\citenamefont {Bogolubsky}\ \emph {et~al.}(2006)\citenamefont
  {Bogolubsky}, \citenamefont {Burgio}, \citenamefont {Mitrjushkin},\ and\
  \citenamefont {M{\"u}ller-Preussker}}]{Bogolubsky:2005wf}%
  \BibitemOpen
  \bibfield  {author} {\bibinfo {author} {\bibfnamefont {I.~L.}\ \bibnamefont
  {Bogolubsky}}, \bibinfo {author} {\bibfnamefont {G.}~\bibnamefont {Burgio}},
  \bibinfo {author} {\bibfnamefont {V.~K.}\ \bibnamefont {Mitrjushkin}}, \ and\
  \bibinfo {author} {\bibfnamefont {M.}~\bibnamefont {M{\"u}ller-Preussker}},\
  }\href@noop {} {\bibfield  {journal} {\bibinfo  {journal} {Phys. Rev.}\
  }\textbf {\bibinfo {volume} {D74}},\ \bibinfo {pages} {034503} (\bibinfo
  {year} {2006})},\ \Eprint {http://arxiv.org/abs/hep-lat/0511056}
  {hep-lat/0511056} \BibitemShut {NoStop}%
%%CITATION = HEP-LAT 0511056;%%
\bibitem [{\citenamefont {Bogolubsky}\ \emph {et~al.}(2008)\citenamefont
  {Bogolubsky}, \citenamefont {Bornyakov}, \citenamefont {Burgio},
  \citenamefont {Ilgenfritz}, \citenamefont {Mitrjushkin},\ and\ \citenamefont
  {M{\"u}ller-Preussker}}]{Bogolubsky:2007bw}%
  \BibitemOpen
  \bibfield  {author} {\bibinfo {author} {\bibfnamefont {I.~L.}\ \bibnamefont
  {Bogolubsky}}, \bibinfo {author} {\bibfnamefont {V.~G.}\ \bibnamefont
  {Bornyakov}}, \bibinfo {author} {\bibfnamefont {G.}~\bibnamefont {Burgio}},
  \bibinfo {author} {\bibfnamefont {E.-M.}\ \bibnamefont {Ilgenfritz}},
  \bibinfo {author} {\bibfnamefont {V.~K.}\ \bibnamefont {Mitrjushkin}}, \ and\
  \bibinfo {author} {\bibfnamefont {M.}~\bibnamefont {M{\"u}ller-Preussker}},\
  }\href {\doibase 10.1103/PhysRevD.77.014504} {\bibfield  {journal} {\bibinfo
  {journal} {Phys. Rev.}\ }\textbf {\bibinfo {volume} {D77}},\ \bibinfo {pages}
  {014504} (\bibinfo {year} {2008})},\ \Eprint {http://arxiv.org/abs/0707.3611}
  {arXiv:0707.3611 [hep-lat]} \BibitemShut {NoStop}%
%%CITATION = 0707.3611;%%
\bibitem [{\citenamefont {Bornyakov}\ \emph {et~al.}(2010)\citenamefont
  {Bornyakov}, \citenamefont {Mitrjushkin},\ and\ \citenamefont
  {M{\"u}ller-Preussker}}]{Bornyakov:2009ug}%
  \BibitemOpen
  \bibfield  {author} {\bibinfo {author} {\bibfnamefont {V.}~\bibnamefont
  {Bornyakov}}, \bibinfo {author} {\bibfnamefont {V.}~\bibnamefont
  {Mitrjushkin}}, \ and\ \bibinfo {author} {\bibfnamefont {M.}~\bibnamefont
  {M{\"u}ller-Preussker}},\ }\href {\doibase 10.1103/PhysRevD.81.054503}
  {\bibfield  {journal} {\bibinfo  {journal} {Phys. Rev.}\ }\textbf {\bibinfo
  {volume} {D81}},\ \bibinfo {pages} {054503} (\bibinfo {year} {2010})},\
  \Eprint {http://arxiv.org/abs/0912.4475} {arXiv:0912.4475 [hep-lat]}
  \BibitemShut {NoStop}%
%%CITATION = 0912.4475;%%
\bibitem [{\citenamefont {Cucchieri}\ \emph {et~al.}(2012)\citenamefont
  {Cucchieri}, \citenamefont {Dudal}, \citenamefont {Mendes},\ and\
  \citenamefont {Vandersickel}}]{Cucchieri:2011ig}%
  \BibitemOpen
  \bibfield  {author} {\bibinfo {author} {\bibfnamefont {A.}~\bibnamefont
  {Cucchieri}}, \bibinfo {author} {\bibfnamefont {D.}~\bibnamefont {Dudal}},
  \bibinfo {author} {\bibfnamefont {T.}~\bibnamefont {Mendes}}, \ and\ \bibinfo
  {author} {\bibfnamefont {N.}~\bibnamefont {Vandersickel}},\ }\href {\doibase
  10.1103/PhysRevD.85.094513} {\bibfield  {journal} {\bibinfo  {journal}
  {Phys.Rev.}\ }\textbf {\bibinfo {volume} {D85}},\ \bibinfo {pages} {094513}
  (\bibinfo {year} {2012})},\ \Eprint {http://arxiv.org/abs/1111.2327}
  {arXiv:1111.2327 [hep-lat]} \BibitemShut {NoStop}%
%%CITATION = ARXIV:1111.2327;%%
\bibitem [{\citenamefont {Fischer}\ \emph {et~al.}(2009)\citenamefont
  {Fischer}, \citenamefont {Maas},\ and\ \citenamefont
  {Pawlowski}}]{Fischer:2008uz}%
  \BibitemOpen
  \bibfield  {author} {\bibinfo {author} {\bibfnamefont {C.~S.}\ \bibnamefont
  {Fischer}}, \bibinfo {author} {\bibfnamefont {A.}~\bibnamefont {Maas}}, \
  and\ \bibinfo {author} {\bibfnamefont {J.~M.}\ \bibnamefont {Pawlowski}},\
  }\href {\doibase 10.1016/j.aop.2009.07.009} {\bibfield  {journal} {\bibinfo
  {journal} {Annals Phys.}\ }\textbf {\bibinfo {volume} {324}},\ \bibinfo
  {pages} {2408} (\bibinfo {year} {2009})},\ \Eprint
  {http://arxiv.org/abs/0810.1987} {arXiv:0810.1987 [hep-ph]} \BibitemShut
  {NoStop}%
%%CITATION = 0810.1987;%%
\bibitem [{\citenamefont {Aouane}\ \emph {et~al.}(2012)\citenamefont {Aouane},
  \citenamefont {Bornyakov}, \citenamefont {Ilgenfritz}, \citenamefont
  {Mitrjushkin}, \citenamefont {M{\"u}ller-Preussker},\ and\ \citenamefont
  {Sternbeck}}]{Aouane:2011fv}%
  \BibitemOpen
  \bibfield  {author} {\bibinfo {author} {\bibfnamefont {R.}~\bibnamefont
  {Aouane}}, \bibinfo {author} {\bibfnamefont {V.~G.}\ \bibnamefont
  {Bornyakov}}, \bibinfo {author} {\bibfnamefont {E.-M.}\ \bibnamefont
  {Ilgenfritz}}, \bibinfo {author} {\bibfnamefont {V.~K.}\ \bibnamefont
  {Mitrjushkin}}, \bibinfo {author} {\bibfnamefont {M.}~\bibnamefont
  {M{\"u}ller-Preussker}}, \ and\ \bibinfo {author} {\bibfnamefont
  {A.}~\bibnamefont {Sternbeck}},\ }\href {\doibase 10.1103/PhysRevD.85.034501}
  {\bibfield  {journal} {\bibinfo  {journal} {Phys.Rev.}\ }\textbf {\bibinfo
  {volume} {D85}},\ \bibinfo {pages} {034501} (\bibinfo {year} {2012})},\
  \Eprint {http://arxiv.org/abs/1108.1735} {arXiv:1108.1735 [hep-lat]}
  \BibitemShut {NoStop}%
%%CITATION = ARXIV:1108.1735;%%
\bibitem [{\citenamefont {Bogolubsky}\ \emph {et~al.}(2007)\citenamefont
  {Bogolubsky}, \citenamefont {Ilgenfritz}, \citenamefont
  {M{\"u}ller-Preussker},\ and\ \citenamefont {Sternbeck}}]{Bogolubsky:2007ud}%
  \BibitemOpen
  \bibfield  {author} {\bibinfo {author} {\bibfnamefont {I.~L.}\ \bibnamefont
  {Bogolubsky}}, \bibinfo {author} {\bibfnamefont {E.-M.}\ \bibnamefont
  {Ilgenfritz}}, \bibinfo {author} {\bibfnamefont {M.}~\bibnamefont
  {M{\"u}ller-Preussker}}, \ and\ \bibinfo {author} {\bibfnamefont
  {A.}~\bibnamefont {Sternbeck}},\ }\href@noop {} {\bibfield  {journal}
  {\bibinfo  {journal} {PoS}\ }\textbf {\bibinfo {volume} {LAT2007}},\ \bibinfo
  {pages} {290} (\bibinfo {year} {2007})},\ \Eprint
  {http://arxiv.org/abs/0710.1968} {arXiv:0710.1968 [hep-lat]} \BibitemShut
  {NoStop}%
%%CITATION = 0710.1968;%%
\bibitem [{\citenamefont {Bogolubsky}\ \emph
  {et~al.}(2009{\natexlab{b}})\citenamefont {Bogolubsky}, \citenamefont
  {Ilgenfritz}, \citenamefont {M{\"u}ller-Preussker},\ and\ \citenamefont
  {Sternbeck}}]{Bogolubsky:2009dc}%
  \BibitemOpen
  \bibfield  {author} {\bibinfo {author} {\bibfnamefont {I.~L.}\ \bibnamefont
  {Bogolubsky}}, \bibinfo {author} {\bibfnamefont {E.-M.}\ \bibnamefont
  {Ilgenfritz}}, \bibinfo {author} {\bibfnamefont {M.}~\bibnamefont
  {M{\"u}ller-Preussker}}, \ and\ \bibinfo {author} {\bibfnamefont
  {A.}~\bibnamefont {Sternbeck}},\ }\href {\doibase
  10.1016/j.physletb.2009.04.076} {\bibfield  {journal} {\bibinfo  {journal}
  {Phys. Lett.}\ }\textbf {\bibinfo {volume} {B676}},\ \bibinfo {pages} {69}
  (\bibinfo {year} {2009}{\natexlab{b}})},\ \Eprint
  {http://arxiv.org/abs/0901.0736} {arXiv:0901.0736 [hep-lat]} \BibitemShut
  {NoStop}%
%%CITATION = 0901.0736;%%
\bibitem [{\citenamefont {von Smekal}\ \emph {et~al.}(2009)\citenamefont {von
  Smekal}, \citenamefont {Maltman},\ and\ \citenamefont
  {Sternbeck}}]{vonSmekal:2009ae}%
  \BibitemOpen
  \bibfield  {author} {\bibinfo {author} {\bibfnamefont {L.}~\bibnamefont {von
  Smekal}}, \bibinfo {author} {\bibfnamefont {K.}~\bibnamefont {Maltman}}, \
  and\ \bibinfo {author} {\bibfnamefont {A.}~\bibnamefont {Sternbeck}},\ }\href
  {\doibase 10.1016/j.physletb.2009.10.030} {\bibfield  {journal} {\bibinfo
  {journal} {Phys. Lett.}\ }\textbf {\bibinfo {volume} {B681}},\ \bibinfo
  {pages} {336} (\bibinfo {year} {2009})},\ \Eprint
  {http://arxiv.org/abs/0903.1696} {arXiv:0903.1696 [hep-ph]} \BibitemShut
  {NoStop}%
%%CITATION = 0903.1696;%%
\end{thebibliography}
%

\end{document}